\documentstyle[prb,aps,epsf]{revtex}
\begin{document}

\twocolumn[\hsize\textwidth\columnwidth\hsize\csname@twocolumnfalse%
\endcsname

\title{Novel massless phase of Haldane-gap antiferromagnets in magnetic field}

\author{G\'abor F\'ath\cite{byline1} and Peter B.\ Littlewood}
\address{Cavendish Laboratory, University of Cambridge \\
Madingley Road, Cambridge CB3 0HE, England}

\date{\today}

\maketitle

\begin{abstract}
The behavior of Haldane-gap antiferromagnets in strong magnetic field is
not universal. While the low-energy physics of the conventional 1D spin-1 
Heisenberg model in its magnetized regime is described by one incommensurate
soft mode, other systems with somewhat perturbed coupling constants can 
possess two characteristic soft modes in a certain range
of the field strength. Such a {\em two}-component Lutinger liquid phase is 
realised above the massive Haldane-gap phase, and in general
above any massive nonmagnetic phase, when the 
ground state exhibits short range incommensurate fluctuations
already in the absence of the field. 
\end{abstract}
\pacs{PACS numbers: 75.10.Jm, 75.40.Cx, 75.10.-b}

\vskip 0.3 truein
]

Quantum mechanical systems possessing a spectral gap in their ground state are
usually very robust: the gap can persist even if relatively large perturbations 
are added to the Hamiltonian. This is exactly the case for the one-dimensional
spin-1 antiferromagnetic chain where the existence of an energy gap, the 
Haldane gap,\cite{Hal} is well documented and understood for
the Heisenberg model. In the more general SU(2) symmetric bilinear-biquadratic
spin-1 model 
\begin{eqnarray}
   H =&& \sum_{i=1}^{N-1} \left[
         {\rm\bf S}_i {\rm\bf S}_{i+1}
         + \beta ({\rm\bf S}_i {\rm\bf S}_{i+1})^2\right]
         -h \sum_{i=1}^N S_i^z.
   \label{spin-1}
\end{eqnarray}
in which $\beta=0$ yields the conventional Heisenberg model,
the Haldane gap survives in the whole range $-1<\beta<1$, thus there is a large 
region in the $\beta$-space where the biquadratic term has seemingly no
effect on the low-energy physics.\cite{Aff} The Haldane gap disappears in the 
integrable critical points $\beta=\pm 1$ beyond which new phases appear.
For $\beta<-1$ translation
invariance is spontaneously broken and dimerization occurs. In the region
$\beta>1$ the bilinear-biquadratic model is believed to remain gapless
with soft modes at momenta $k=0,\pm 2\pi/3$. Since the system is one 
dimensional no long-range order exists in this phase either: 
the correlation functions decay algebraically.

The emergence of the Haldane gap in the vicinity of the $\beta=1$ point,
the so called Uimin-Lai-Sutherland (ULS) point,\cite{Uim-Lai-Sut} has been 
investigated recently by Itoi and Kato\cite{Ito-Kat} in the absence of magnetic
fields $h=0$. The ULS model has an SU(3) symmetry which is broken down to 
SU(2) when $\beta$ is tuned away from 1. They identified the
critical theory of the ULS point with the $k=1$ SU(3) 
Wess-Zumino-Witten-Novikov (WZWN) model and concluded that the SU(3) symmetry
breaking perturbation, represented by the deviation term with $(\beta-1)$
in the Hamiltonian,
is irrelevant for $\beta>1$, i.e., the system remains critical there, but it
becomes marginally relevant and gives rise to a dynamic mass generation 
(the Haldane gap) for $\beta<1$. The transition is of the 
Berezinskii-Kosterlitz-Thouless (BKT) type, i.e., the gap open exponentially 
slowly away from the ULS point. These findings are supported by earlier 
numerical results.\cite{Fat-Sol}

When there is a magnetic field present the SU(2) symmetry of the
bilinear-biquadratic model breaks down to U(1). 
At the ULS point, however, where the
symmetry is higher, the quantities $N_+,N_0$, and $N_-$, denoting
the numbers of $+$, $0$, and $-$ spin states in the wave function, with 
$N_+ +N_0 +N_-=N$ the length of the chain, are independently conserved.
Thus the remaining continuous symmetry for $\beta=1$, $h>0$ is 
$U(1)\times U(1)$. For $\beta<1$ when the strength of the field is higher 
then the value of the gap, the Haldane
gap collapses and the low-energy physics of the magnetized system is 
governed by gapless excitations. The emerging periodicity is a function
of the field strength and in the generic situation incommensurate. Finally, when
the field is strong enough all the spins align and the magnetization
saturates.

At the ULS point, where the Haldane gap no longer exists, 
the analysis of the Bethe Ansatz equations showed\cite{Par-Kiw} that 
the magnetization growth is not smooth.
There is a second order phase transition which leads to a cusp
in the the magnetization curve $m=m(h)$ at a critical field 
$h_c\approx 0.941$, $m_c=m(h_c)\approx 0.556$.
In the low field regime $h<h_c$, all three
probabilities $P_+=N_+/N$, $P_0=N_0/N$ and $P_-=N_-/N$ tend to finite values
as $N\to\infty$, but for $h>h_c$ the ground state is
in a sector with $N_-=0$, i.e., $P_-=0$. Thus $P_-$ can be
used as an order parameter for the transition.
The magnetization behaves continuously at the
critical point, but the susceptibility diverges below $h_c$.
The high-field phase (S1 phase), where the 
low-energy sector is identical to that of the spin-1/2 Heisenberg chain,
can be described by a one-component Luttinger liquid, and there is only one 
characteristic soft mode. The low field phase (S2 phase), on the other hand, is 
equivalent to a two-component Luttinger liquid, possessing two critical
degrees of freedom, and thus
{\em two} incommensurate soft modes. Their positions
in the Brillouin zone are functions of the probabilities $P_{+,0,-}$.
In the sequel, we label phases
by S$n$, where $n=0,1,2$ stands for the number of critical, Luttinger
liquid components. Note that S0 is a gapped phase.

Early speculations\cite{Par-Kiw,Par-Bon} that the S2-S1 phase transition 
of the ULS model may also take place in the Heisenberg 
chain at $\beta=0$ was finally refuted by Takahashi and Sakai\cite{Tak-Sak} 
who found that in the
whole range $0<m<1$ the low-energy physics is described by a $c=1$ U(1) 
conformal field theory (CFT) which is equivalent to the one-component Luttinger
liquid, thus only an S1 phase appears.
The Luttinger liquid parameters vary smoothly as a function of the
magnetization.
Naturally arises the question whether the appearence of the Haldane gap in the 
ground state only allows the S1 behavior seen for the pure 
Heisenberg model, or there is a certain domain in parameter space 
where multi-component Luttinger liquids such as an S2 phase can occur above the
S0 Haldane phase. To clarify this question is the principal aim of this Letter. 

The first indication that the Haldane gap of the bilinear-biquadratic
model may collapse into an S2 phase
comes from the numerical observation that at $h=0$ 
the VBS point $\beta_{\rm VBS}=1/3$, where the ground state
can be constructed exactly using nearest-neighbor valence bonds, is in fact
a disorder point, beyond which short range fluctuations in the ground
state become incommensurate.\cite{Bur-Sch} However, due to the finite 
correlation length, the peak at $\pi$ in the static structure factor only 
splits somewhat later at the so called Lifshitz point $\beta_{\rm Lifs}\approx 
0.438$.\cite{Bur-Sch} One can define a third special point 
$\beta_{\rm Disp}$, which is {\em a priori} distinct from (but close to)
the above two, where the emerging incommensurability make the position of the 
minimum gap in the energy-momentum dispersion relation move away from $\pi$.
In the range $\beta_{\rm Disp}<\beta<1$ the momentum $p_{\rm H}$
associated with this gap minimum rapidly shifts from the antiferromagnetic
value $\pi$ to the ULS value $2\pi/3$. When the magnetic field
reaches the value of the gap at $p_{\rm H}$, and the system starts 
to become magnetized, there are obviously two characteristic momenta in
the system: one is $2\pi m$, as suggested by the generalization of the
Lieb-Schultz-Mattis theorem,\cite{Osh-Aff} and the other is the finite 
difference of the two split gap minima $2p_{\rm H}$. Of course, any linear
combinations of these two may also show up in the correlation functions, 
which will be dominated by the most slowly decaying terms, 
or on shorter distances, by the ones having the largest amplitude. 
 
A more quantitative analysis is possible on the basis of the Bethe
Ansatz (BA) solution of the ULS model, when the deviation term 
proportional to $(1-\beta)$ in the Hamiltonian is treated as a perturbation.
The ULS model is solvable by the two-component 
nested BA method.\cite{Uim-Lai-Sut}
This associates the $+$ spin
states with an inert background in which particles with two possible internal
states, spin $0$ and $-$, move. Two sets of spectral parameters
are introduced, one for the particles (component-1) and one for their internal
state (component-2). Their actual values can be calculated
by solving a set of coupled algebraic (or in the $N\to\infty$ limit integral)
equations. The BA technique also allows one to obtain
the finite size corrections ${\cal O}(1/N)$
to the low-energy excitations near the thermodynamic limit. In the S2 phase
of the ULS model the energy spectrum has the following structure:\cite{Ize-Fra}
\begin{eqnarray}
   \delta E=\! E_{\bf a}-\! E_g &=&\!
   \frac{2\pi}{N}\left[ v_1(\Delta_1^+ +\Delta_1^- )+
                v_2(\Delta_2^+ +\Delta_2^-) \right]
   \label{E_diff}
\end{eqnarray}
with
\begin{eqnarray}
   \Delta_1^\pm &=& {1\over 2} \left[Z_{11}d_1+Z_{21}d_2 \pm
                  \frac{Z_{22}l_1 -Z_{12}l_2}{2\det Z} \right]^2
                  + n_1^\pm     \nonumber\\
   \Delta_2^\pm &=& {1\over 2} \left[Z_{12}d_1+Z_{22}d_2 \mp
                  \frac{Z_{21}l_1 -Z_{11}l_2}{2\det Z} \right]^2
                  + n_2^\pm
   \label{Deltas}
\end{eqnarray}
where $E_g$ is the energy of the ground state, the index ${\bf a}\equiv
\{d_1,d_2;l_1,l_2;n_1^+,n_1^-,n_2^+,n_2^-\}$ is a shorthand for eight
integer (half-integer) quantum numbers specifying the eigenstate, 
and the matrix $Z_{\alpha\beta}$, $\alpha,\beta=1,2$ is the ''dressed
charge matrix`` responsible for the interaction of the two BA components.
The relative momentum of the state ${\bf a}$ reads
\begin{eqnarray}
   \delta P= P_{\bf a}-P_g = Q +
       \frac{2\pi}{N}(&&\Delta_1^+ -\Delta_1^- +\Delta_2^+-\Delta_2^-) 
   \label{P_diff}
\end{eqnarray}
where $Q=Q_{\bf a}$ is an ${\cal O}(1)$ term 
\begin{eqnarray}
   Q = &&2\pi(1-P_+) d_1+ 2\pi P_- d_2 +\pi l_1.
   \label{eq:Q}
\end{eqnarray}
The physical interpretation of the quantum numbers is as follows:
$l_\alpha$ ($d_\alpha$) represents the number of particles added to 
(transferred from the left Fermi point to the right in) component $\alpha$.
$n^\pm_\alpha$ is the number of small momentum particle-hole pairs 
created in component $\alpha$ around the left ($-$) and the right ($+$) 
Fermi points. While $l_\alpha$ and $n^\pm_\alpha\ge 0$ are always
integers, $d_\alpha$ is integer or half integer with $d_{1,2}\equiv l_{2,1}/2 
\; ({\rm mod}\; 1)$.\cite{Sch-Mut-Kar} 
This structure is analogous to the one found in the 1D Hubbard model where
the two components are called "charge" and "spin", resp. In the present 
case $l_1=\delta N_0+\delta N_-$, $l_2=\delta N_-$ for which there
are selection rules when a given type of correlation functions is considered. 

\begin{figure}[hbt]
\epsfxsize=\columnwidth\epsfbox{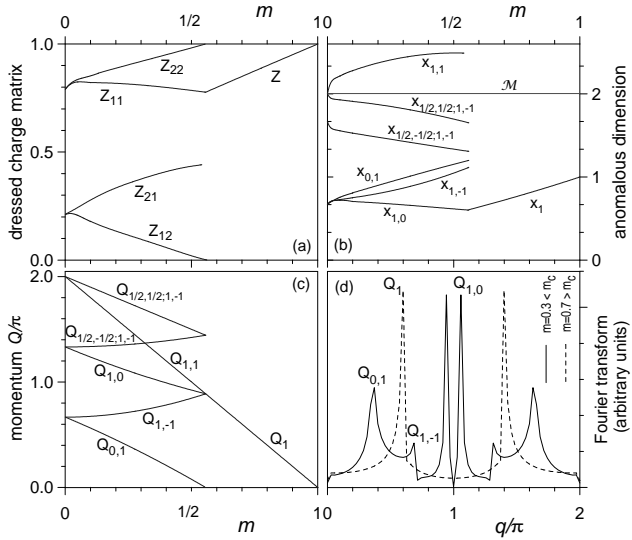} 
\caption{Bethe Ansatz results for the ULS model vs the magnetization $m$.
(a) The dressed charge matrix, (b) anomalous dimension and (c) momentum
$Q$ of some selected operators. Some other operators marginal at $h=0$ are
not shown. (d) DMRG results for the Fourier transform of the one-point
function $\langle S_n^z\rangle$. The peak at the Lieb-Schultz-Mattis value 
$q=2\pi m, 2\pi (1-m)$ only develops for $m>m_c$.} 
\label{fig:ULS} 
\end{figure}

The low-energy excitations of the ULS model in its S2 phase can be interpreted
by assuming that they arise from the direct sum of two $c=1$ conformal
field theories (CFT) each having a different sound velocity $v_1$ and $v_2$, 
resp.\cite{Ize-Fra} As is indicated by Eq.\ (\ref{Deltas}) 
local physical operators necessarily couple to both CFTs. 
Conformal invariance then requires that the 
2-point functions behave as\cite{Ize-Fra}
\begin{eqnarray}
   &&\langle \phi(x,t)\phi(0,0) \rangle = 
   A_\phi e^{-iQx} \prod_{\alpha,\pm} (x\mp iv_\alpha t)^{-2\Delta_\alpha^\pm}
   \label{corr}
\end{eqnarray}
showing the analog of "spin-charge separation" for the present spin-1 situation.
Let us consider the operator content of the theory: each operator 
$\phi_{\bf a}$ (primary or secondary) is labeled by the eight quantum 
numbers ${\bf a}$, and has the anomalous dimension $x_{\bf a}=
\Delta_1^+ +\Delta_1^- +\Delta_2^+ +\Delta_2^-$ and conformal spin
$s_{\bf a}=\Delta_1^+ -\Delta_1^- +\Delta_2^+ -\Delta_2^- =
d_1 l_1+d_2 l_2+n_1^+ -n_1^- +n_2^+ -n_2^-$. Note that the total momentum
associated to these operators is $\delta P$ in Eq.\ (\ref{P_diff}),
involving the $Q$ term as well. 
There are four marginal operators ${\cal M}_{1,2,3,4}$
which can be formed using only the $n_{1,2}^\pm$ quantum numbers and
setting $l_\alpha,d_\alpha=0$. Their dimension, spin and total momentum 
($x=2,\; s=0,\;\delta P=0$) 
do not depend on the $Z_{\alpha\beta}$ matrix. The presence of these 
operators causes the existence of an extended critical domain in the 
parameter space.
The other relevant or marginal operators are all primary, i.e., 
$n_\alpha^\pm=0$ and all depend on the $Z_{\alpha\beta}$ matrix.
This latter can be calculated numerically for the ULS model by solving a set
of coupled integral equations. The results\cite{Fat-San} are shown in 
Fig.\ \ref{fig:ULS}(a). When $h=0$, $Z_{11}=Z_{22}=\sqrt{1/3+1/2\sqrt{3}}$, 
$Z_{12}=Z_{21}=\sqrt{1/3-1/2\sqrt{3}}$, and there is an additional marginal
operator $\phi_{1/2,1/2;1,-1}$ (and its equivalents under SU(3) and 
parity transformations, e.g., $\phi_{1,1}$) as can be checked 
using Eq.\ (\ref{Deltas}). (From now on in the index ${\bf a}$ we omit 
$l_\alpha$ and/or $n_\alpha^\pm$ when they are zero.)
The operator $\phi_{1/2,1/2;1,-1}$ has anomalous dimension 
$x=2$ and total momentum $\delta P=0$ when $h=0$. 
This is the principal operator responsible for the SU(3)$\to$SU(2) symmetry
breaking processes $00\longleftrightarrow +-,-+$.
As was shown in Ref.\ \cite{Ito-Kat} the interplay of $\phi_{1/2,1/2;1,-1}$
and the ${\cal M}$ operators (which constitute the SU(3) current
interaction in the WZWN theory)
gives rise to the Haldane gap for $\beta<1$ but maintains criticality for
$\beta>1$. 

When $0<h<h_c$ the anomalous dimensions and momenta of the operators present
in the ULS model vary as shown in Fig.\ \ref{fig:ULS}(b). Since the 
perturbation part of the Hamiltonian, represented by the deviation $(1-\beta)$
term, transforms under translations with momentum zero, in its
decomposition into the operators present in the ULS model at $h>0$
only operators with $\delta P=0$ can appear. This is a serious limitation,
since as shown in Fig.\ \ref{fig:ULS}(c), the two characteristic
momenta $Q_{1,0}=2\pi(1-P_+)$ and $Q_{0,1}=2\pi P_-$, associated to
the large momentum transfer processes of the two components in Eq.\ 
(\ref{eq:Q}), become generically incommensurate.
$\phi_{1/2,1/2;1,-1}$ has no longer $\delta P=0$ so it does not
contribute. It is in fact an Umklapp operator which appears in the
low-energy description only at $h=0$. 
What contribute are the operators $\phi_{d_1,d_2;1,-1;n_1^\pm;n_2^\pm}$
with $d_1,d_2$ half-integer, and $n_\alpha^\pm>0$ 
chosen in a way to reinstall
$\delta P=0$. However, due to the appearance of the necessary small momentum
particle-hole excitations such operators are highly irrelevant.
The marginal operators ${\cal M}$ are not able alone to drive the two
Luttinger liquid components
away from criticality, although they make the universality class change 
continuously. We conclude that when $0<m<m_c$ the bilinear-biquadratic model
must remain in its S2 phase in an extended domain on {\em both} sides of the 
ULS line $\beta=1$. 

\begin{figure}[hbt]
\epsfxsize=\columnwidth\epsfbox{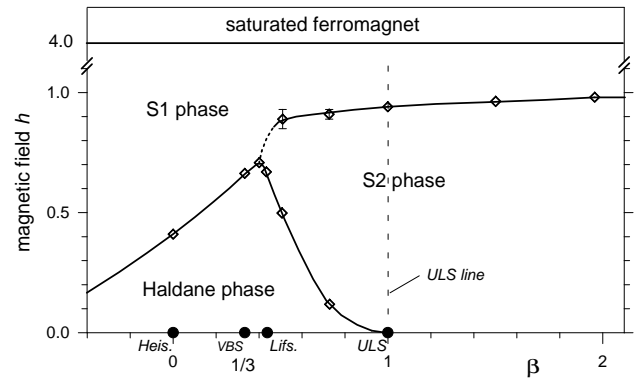}  
\caption{Phase diagram of the bilinear-biquadratic spin-1 chain in a
magnetic field. The one- and two-component Luttinger liquid phases are
denoted by S1 and S2, resp. The S1--S2 phase boundary is determined by the
DMRG ($\Diamond$ points). The dotted line indicates some uncertainties near
the disorder points.} 
\label{fig:ph_diag} 
\end{figure}

When the magnetic field reaches the critical value $h_c$ in the ULS
model the sound velocity $v_2\to 0$, and the corresponding critical
degree of freedom becomes massive. The emerging S1 phase
can be described by a single $c=1$ CFT, and the universality class is
determined by a scalar dressed charge $Z$. Once again we do not expect any
drastical changes in the low-energy physics as $\beta$ is perturbed away
from 1. The predicted phase diagram is shown in Fig.\ \ref{fig:ph_diag}.

In order to find the exact phase boundaries we carried out a detailed
numerical analysis using the DMRG technique. Unfortunately the DMRG does
not work in momentum space thus a direct search for tracking soft modes
by calculating energy gaps is not feasible. Instead, we calculated
the decay of the one-point function $\langle S_j^z\rangle$ 
from the edge of an open chain for different
values of $\beta$ and magnetization $m=S^z_{\rm tot}/N$. The one-point
function contain the same information as the equal time two-point 
correlation function in Eq.\ (\ref{corr}) (note that the exponent of 
the one-point function
is half of that of the two-point function), and the appearing soft modes
can be identified in the Fourier transforms or by making a suitable
multi-parameter fit in real space.\cite{Fat-San}

Once again exact results can be obtained for the ULS model. Considering 
$\langle S_j^z\rangle$ in the S2 phase only operators with $\delta
S_{\rm tot}^z=0$, i.e., $l_1=l_2=0$ and $l_1=-l_2=\pm 1$ can contribute.  
The most relevant operators are $\phi_{1,0}$, $\phi_{0,1}$, $\phi_{1,-1}$,
$\phi_{1/2,1/2;1,-1}$, and $\phi_{1/2,-1/2;1,-1}$
(and their equivalents under symmetries). The associated momenta $Q$
(position of the peak in the structure factor) and the occurring critical
exponents can be read off from Fig.\ \ref{fig:ULS}(b) and (c). 
The example presented in Fig.\ \ref{fig:ULS}(d) illustrates that in the
S2 phase of the ULS model only the first three with $l_1=l_2=0$ have
nonvanishing amplitude as dictated by the higher symmetry. However,
$\phi_{1/2,\pm 1/2;1,-1}$ contributes when $\beta\ne 1$.
It is remarkable that in the S2 phase the operator $\phi_{1,1}$,
which has an anomalous dimension over 2 and momentum $Q_{1,1}=2\pi(1-m)$, 
does not contribute to the correlation function. In the high-field S1
phase, however, the only peak discernible in the structure factor, as
shown in Fig.\ \ref{fig:ULS}(d), is the
one with $Q_1=2\pi(1-m)$, in agreement with the Lieb-Schultz-Mattis 
theorem.\cite{Osh-Aff} The presence or absence of a peak in the
structure factor at $k=\pm 2\pi m$ can thus be used efficiently 
to distinguish between the two phases and locate the phase boundary.
Alternatively, one can monitor the amplitude of the peak at $Q_{0,1}$
which tends to zero on the transition line where the multi-peak
structure collapses into a single-peak one. The phase diagram, as
determined by the DMRG calculation, is shown in Fig.\ \ref{fig:ph_diag}.
Details of the numerical investigation will be published 
elsewhere.\cite{Fat-San}

Our analysis is valid in the strict sense close to the ULS line only.
Here the S2--S1 phase transition is clearly associated to the
depletion of one of the two "bands`` as suggested by the BA. Note that in
Fig.\ \ref{fig:ph_diag} the phase boundary does not change very much
until about $\beta\sim 0.5$ where it seems to decline rather
rapidly. {\em A priori} we cannot exclude the possibility that some 
operators become relevant here and open a gap in one of the critical
components. This question needs further clarification.

In the bilinear-biquadratic model in Eq.\ (\ref{spin-1}) the S2 phase
terminates near $\beta\approx 0.4$. This is still far in the parameter space
from the currently known spin-1 Haldane gap materials for which the 
biquadratic term is small, and the pure Heisenberg model, although with
some anisotropies, is a good description. For these systems only S1 type
massless phases (and eventually, for some special values of $m$, additional S0 type
phases, i.e., magnetization plateaus\cite{Osh-Aff,Tot}) are expected to appear. 
However, even here, due to the closeness of the S2 phase somewhat further
in the phase diagram, massive but relatively low energy excitations are
predicted to show up in the weakly magnetized regime. They are expected to
contribute in experiments probing higher lying excitations such as 
in inelastic neutron scattering, or in
situations when short distance physics is important as, e.g., in 
nonmagnetically doped materials. Although there is no sharp phase
transition in this case, the low-field and high-field regimes may look rather
different, separated by a more or less narrow crossover region as
observed, e.g., in Ref.\ 7. 


In general, a two-component Luttinger liquid phase (S2), and then an eventual
phase transition S2$\to$S1 
during the magnetization process, is expected to occur whenever the ground 
state develops incommensurate fluctuations already at $h=0$. 
It must not necessarily be above a Haldane gap;
the spin-1/2 zig-zag ladder, for example, which is expected to describe 
adequately the quasi-1D antiferromagnet Cs$_2$CuCl$_4$, where the gap 
is due to dimerization and fluctuations are also predicted to be 
incommensurate already without a magnetic field,\cite{Ner-Gog-Ess} is
another possible candidate.

Enlightening discussions with P.\ Santini, A.\ Schmitt and A.\ Tsvelik are
appreciated. This research was supported by the EPSRC grant no.\ GR/L55346.

\end{document}